\documentclass[12pt,a4,epsf]{article}
\global\arraycolsep=2pt 
\input{epsf}
\usepackage{graphicx}
\begin{document}
\makeatletter
\def\fmslash{\@ifnextchar[{\fmsl@sh}{\fmsl@sh[0mu]}}
\def\fmsl@sh[#1]#2{%
  \mathchoice
    {\@fmsl@sh\displaystyle{#1}{#2}}%
    {\@fmsl@sh\textstyle{#1}{#2}}%
    {\@fmsl@sh\scriptstyle{#1}{#2}}%
    {\@fmsl@sh\scriptscriptstyle{#1}{#2}}}
\def\@fmsl@sh#1#2#3{\m@th\ooalign{$\hfil#1\mkern#2/\hfil$\crcr$#1#3$}}
\makeatother
\thispagestyle{empty}
\begin{titlepage}
\begin{flushright}
hep-ph/0206091 \\
LMU 02/05 \\
\today
\end{flushright}

\vspace{0.3cm}
\boldmath
\begin{center}
  \Large {\bf Seesaw induced Higgs Mechanism}
\end{center}
\unboldmath
\vspace{0.8cm}

\unboldmath
\vspace{0.8cm}
\begin{center}
  {\large X. Calmet \footnote{supported by the Deutsche
Forschungsgemeinschaft, DFG-No. FR 412/27-2, \\
email:calmet@theorie.physik.uni-muenchen.de}}\\
  \end{center}
  \vspace{.3cm}
\begin{center}
{\sl Ludwig-Maximilians-University Munich, Sektion Physik}\\
{\sl Theresienstra{\ss}e 37, D-80333 Munich, Germany}
\end{center}
\vspace{\fill}
\begin{abstract}
\noindent
 We discuss a two scalar doublets model which induces the Higgs
  mechanism by means of a seesaw mechanism. This model naturally
  predicts a light Higgs scalar whose mass is suppressed by the grand
  unification scale. The model predicts an intermediate scale between
  the electroweak symmetry breaking scale and the grand unification
  scale at $10^9$ GeV. Below this intermediate energy scale the usual
  standard model appears as an effective theory. A seesaw mechanism in
  the scalar sector of the model not only induces the standard Higgs
  mechanism, but also solves the hierarchy problem. An implementation
  of this mechanism in models where the Planck scale is in the TeV
  region is discussed.
\end{abstract}  
\end{titlepage}
The electroweak symmetry breaking in the standard model
\cite{Glashow}, which is a gauge theory based on the structure group
$SU(3) \times SU(2) \times U(1)$, is implemented by means of the Higgs
mechanism \cite{Higgs:1964pj}.  If the standard model is embedded into
a grand unified theory like e.g. $SU(5)$ \cite{Georgi:1974sy} or
$SO(10)$ \cite{Fritzsch:1975nn}, it suffers from two problems that
have often been discussed in the literature. The running of the
coupling constant of the standard model suggests that the unification
is taking place at a grand unification scale $\Lambda_{GUT}\approx
10^{16}$ GeV.  The gauge hierarchy problem states that it is unnatural
for the electroweak breaking scale $\Lambda_{EW}\approx 174$ GeV to be
so small compared to the fundamental scale $\Lambda_{GUT}$.

A second potential problem with the standard model is that the Higgs
boson is an elementary scalar field. It is useful, in order to get some
intuition, to regularize the theory using a cutoff $\Lambda$.  One
finds that the Higgs squared boson mass $m_H^2$ receives quadratic
``corrections''
\begin{eqnarray}
 m_H^2 &\approx& {m_H^{0}}^2 + \\ \nonumber && +
  \frac{3 g^2 \Lambda^2}{32 \pi^2 m_W^2}
  \left(  m^2_H + 2 m^2_W + m^2_Z   - 
    4  \sum_f  \frac{n_f}{3}
    m_f^2  \right)
\end{eqnarray}
where $g$ is the $SU(2)$ gauge coupling, $m_W$ and $m_Z$ are the
masses of the electroweak gauge bosons, $m_f$ stands for a fermion
mass and where the sum runs over the fermion flavors. The standard
model is a renormalizable theory \cite{'tHooft:rn}. However, if the
intuitive cutoff $\Lambda$ which is used to regularize the theory, is
identified with the grand unification scale $\Lambda_{GUT}$, it seems
to require an unnatural adjustment to keep the Higgs boson mass small
compared to the scale $\Lambda_{GUT}$.  This is the naturalness
problem \cite{'tHooft:1980xbis}.

Besides these problems, the standard model has another unpleasant
feature, the sign of the Higgs boson doublet squared mass has to be
chosen to be negative to trigger the Higgs mechanism. In other words
the Higgs boson doublet is a tachyon. This may be the sign that some
mechanism triggering this phase transition is missing.

Different scenarios have been proposed to cure these problems.  A
dynamical effect could be at the origin of this phase transition, see
ref.  \cite{Hill:2002ap} for a review. In that case it is not
necessary to introduce elementary bosons in the model.  A
supersymmetric extension of the standard model \cite{Haber:1984rc} is
also conceivable. But, in that case the problem of breaking the gauge
symmetry is shifted to that of supersymmetry breaking which remains
unsolved. Other scenarios that are solving the hierarchy problem by
lowering the Planck scale and possibly also the scale for grand
unification to the TeV region have been proposed
\cite{Arkani-Hamed:1998rs}.

In this Letter, we present a minimal extension of the standard model
which is able to address the hierarchy problem and which can be seen
as a limit case of a two Higgs doublets model \cite{Atwood:1996vj}. We
consider the same action as that of the standard model but with a
modified scalar sector.  The first scalar boson is denoted by $h$ and
the second scalar doublet by $H$.  Both doublets have exactly the same
quantum numbers as the usual standard model Higgs doublet. The Yukawa
sector involves only the boson $h$.  The scalar potential is chosen
according to
\begin{eqnarray} \label{action1}
S_{scalar}&=& - \int d^4x \left( h^\dagger  H^\dagger \right )
\left(\begin{array}{cc} 0 & m^2 \\ m^2 & M^2 \end{array}
\right )
\left(\begin{array}{c} h  \\ H \end{array}
\right ) - \\ \nonumber &&  - \int d^4x \lambda_h (h^\dagger h)^2
-  \int d^4x \lambda_H
(H^\dagger H)^2
\end{eqnarray}
i.e. we assume that the first boson $h$ is massless whereas the second
boson $H$ is massive. At this stage the electroweak gauge symmetry is
still unbroken. Taking radiative corrections into account, one expects
that the boson $h$ will get a small, possibly negative, squared mass
according to the Coleman-Weinberg mechanism \cite{Coleman:jx}.  This
mechanism yields a mass for the first scalar boson of the order of 10
GeV \cite{Gunion:1989we}.  In the Coleman-Weinberg case, the scalar
boson mass is a calculable quantity and turns out to be small, one
could thus argue that in that particular case, i.e. when the tree
level mass of the scalar boson is vanishing, one naturally
obtains a light scalar boson mass when radiative corrections are taken
into account. Nevertheless, this calculation involves a
renormalization procedure and thus does not solve the naturalness
problem as it usually formulated.  As we shall see, the mass obtained by
the Coleman-Weinberg mechanism for the first boson $h$ is small
compared to the two other scales $m$ and $M$ involved in the model and
this contribution can be neglected.  On the other hand, we shall
assume that the mass of the second boson $H$ is large and typically of
the order of the fundamental scale of the model. The usual argument is
that the mass of the second boson $H$ receives a large contribution
because its ``naked'' mass is non-vanishing, and if the model is
embedded into a grand unified theory, a natural mass scale for the
mass of the boson $H$ is the grand unification scale.  One thus
expects $M\sim \Lambda_{GUT}$.  In that case we have a large hierarchy
and it is difficult to understand why the scale of the electroweak
interactions is so small compared to $\Lambda_{GUT}$. On the other
side, the situation is quite similar to the situation in neutrino
physics where a large Majorana scale is used to explain a small
neutrino mass, see ref.  \cite{Fritzsch:1999ee} for reviews.  Such a
situation also appears in flavor physics where the light quark masses
are solely induced by mixing with heavy quarks
\cite{Fritzsch:1979zq}. If the parameter $m$ is not too large and not
too small, a seesaw mechanism \cite{Gell-Mann:vs} can be applied. We
shall discuss values of $m$ for which this mechanism can be applied
later.  The seesaw mechanism has been applied in top-color
condensation models \cite{Dobrescu:1997nm} to generate the electroweak
symmetry breaking and it has also recently been applied to a
supersymmetric model in an attempt to solve the so-called
$\mu$-problem \cite{Ito:2000cj}.

After diagonalization of the mass matrix in eq. (\ref{action1}) using
\begin{eqnarray} \label{rotation}
R=\left(\begin{array}{cc} 1& \frac{m^2}{M^2}
    \\  -\frac{m^2}{M^2} & 1 \end{array}
\right ) \approx \left(\begin{array}{cc} 1& 0
    \\ 0 & 1 \end{array}
\right ),
\end{eqnarray}
we obtain the squared masses of the mass eigenstates
\begin{eqnarray} \label{massmatrix}
{\cal M}^2 \approx
\left(\begin{array}{cc} - \frac{m^4}{M^2} & 0 \\ 0 & M^2 \end{array}
\right ).
\end{eqnarray}

The first boson $h$ has become a Higgs boson with a negative squared
mass given by
\begin{eqnarray}
m_h^2=- \frac{m^4}{M^2}
\end{eqnarray}
whereas the second scalar boson $H$ has a positive squared mass of the
order of $\Lambda_{GUT}^2$ and is thus not contributing to the
electroweak symmetry breaking. The mass of the physical Higgs boson is
given by

\begin{eqnarray} \label{physhiggsmass}
M^{phys}_h=\sqrt{2}\frac{m^2}{M}.
\end{eqnarray}

One finds that a Higgs boson mass of the order of 100 GeV can be
obtained if $m\sim 10^9$ GeV using $\Lambda_{GUT}\sim 10^{16}$ GeV.
The price to pay to solve the hierarchy problem is to assume the
presence of an intermediate scale at $10^9$ GeV. But, this mass scale
for $m$ is quite natural too. The parameter $m^2$ also receives
quadratic ``corrections'' due to loop diagrams involving the
electroweak bosons. A natural scale for the parameter $m^2$ is
\begin{eqnarray}
\Lambda_{m}^2\sim \Lambda_{CW}  \Lambda_{GUT}
\end{eqnarray}
where $\Lambda_{CW}\approx 80$ GeV is fixed by the mass scale of the
gauge bosons giving rise to the Coleman-Weinberg mass
\cite{Coleman:jx}. The intermediate scale $\Lambda_{m}$ is a natural
one because the loop diagrams connect the heavy scalar boson $H$,
whose typical scale is $\Lambda_{GUT}$, to the light scalar boson $h$,
whose typical scale is the mass scale generated by the
Coleman-Weinberg mechanism.  This is an intuitive argument which is in
the spirit of the seesaw mechanism when applied to neutrinos. In the
neutrino case \cite{Gell-Mann:vs,Fritzsch:1999ee}, the off-diagonal
terms are assumed to be naturally of the order of the electroweak
scale because the Dirac neutrino is a $SU(2)$ doublet. On the other
hand the Majorana neutrino's typical mass is of the order of the grand
unification because it is a $SU(2)$ singlet. Our case is analogous,
the only difference is that we apply the seesaw mechanism to the
squared mass matrix, the off-diagonal terms are then naturally of the
order of the product of the two scales. One finds $\Lambda_{m} \sim
9\times 10^8$ GeV $\sim m$.  This intermediate mass scale is also
natural in $SO(10)$ grand unification models with an intermediate
breaking scale \cite{Mohapatra:1992dx} where it roughly corresponds to
the scale of the breaking of the $SU(4)_C$ Pati-Salam gauge group
\cite{Pati:1974yy}.  Inserting $m \sim 9\times 10^8$ GeV into eq.
(\ref{physhiggsmass}) yields $M^{phys}_h={\cal O}(100$ GeV), which
should not be taken as a calculation of the Higgs boson's mass but
rather as a confirmation that a Higgs boson mass in the $100$ GeV
region is natural.

Another way to stabilize the intermediate scale $\Lambda_m$ would be
to require that the off-diagonals terms, e.g. $m^2 h^\dagger H$, arise
from an interaction of the type $N^\dagger N h^\dagger H$, $N$ being a
new scalar field, which yields a term $v_N^2 h^\dagger H$, where
$v^2_N$ is the vacuum expectation value of the operator $N^\dagger N$.
Note that the boson $N$ does not need to be charged under $SU(2)$.
The scale of the mass of the boson $N$ could be stabilized by another
seesaw-Higgs mechanism with a new intermediate scale between $10^9$
GeV and $\Lambda_{GUT}$. This new scale might again be stabilized by a
new intermediate scale and we can introduce as much scalar doublets as
necessary to insure the stabilization of these mass scales and this till
the grand unification scale is reached.

As in the standard model we can make use of the unitarity gauge to
rotate the first doublet $h$:
\begin{eqnarray}
h &=& \frac{1}{\sqrt{2}} \left ( \matrix{0 \cr  \eta + v } \right)
\end{eqnarray}
where the vacuum expectation value of Higgs boson is given by
\begin{eqnarray}
v=\sqrt{\frac{m^4}{M^2 \lambda_h}}=\frac{m^2}{M} \frac{1}{\sqrt{\lambda_h}}
\end{eqnarray}
which is naturally a small number. The electroweak scale
$\Lambda_{EW}$ can be defined by $\Lambda_{EW}=v/\sqrt{2}$, and is
thus a small number too.  The hierarchy problem is solved in our
framework.

The masses of the electroweak bosons are given as usually by $m_{W}=g
v/2$ and $m_Z= m_{W} \sqrt{1+g'^2/g^2}$, where $g'$ is the $U(1)$
gauge coupling. The fermion masses are generated by the Yukawa
mechanism as we had assumed a Yukawa type coupling between the doublet
$h$ and the fermions. The four heavy degrees of freedom contained in
$H$ decouple from the remaining of the model and we are left at
energies well below the intermediate scale $m$ with the usual standard
model.  It should be noted that one could relax the assumption
concerning the absence of Yukawa couplings between the heavy scalar
doublet $H$ and the standard model fermions as flavor changing neutral
current transitions are suppressed by the grand unification scale,
note also that the heavy scalar doublet $H$ is not developing a vacuum
expectation value. If the doublet $H$ has Yukawa type couplings, it
leads to point-like four-fermions interactions which are suppressed by
the grand unification scale, once the heavy degrees of freedom have
been integrated out of the theory. It is conceivable, if the Yukawa
couplings are strong enough, that fermion condensates will form
\cite{Bardeen:1989ds}. The electroweak symmetry breaking could be a
combination of two effects: a fundamental scalar boson and a composite
scalar boson could both contribute to the gauge symmetry breaking, in
the spirit of the model proposed in ref.  \cite{Kaplan:1983fs}.

In our framework we can turn to our advantage the fact that
fundamental bosons receive large quadratic corrections or are nearly
massless. The Higgs mechanism appears naturally as a consequence of
the seesaw mechanism, and is triggered by the large gauge hierarchy.
We note that the dual description of the standard model presented in
\cite{Calmet:2000th} allowing a calculation of the weak mixing angle
and of the Higgs boson mass, remains valid in this framework because
the hierarchy between the two mass scales is huge.

In models where the grand unification takes place at a very high scale
of the order of $10^{16}$ GeV, our model will be difficult to
distinguish from the standard model, whereas it might be easier to do
so in models where the unification scale is lower
\cite{Arkani-Hamed:1998rs}.  The diagonalized squared mass matrix
${\cal M}^2$ is actually defined via an expansion in the parameter
$M^2 x$ with $x=\frac{m}{M}$, given by:
\begin{eqnarray}
{\cal M}^2 \approx
\left(\begin{array}{cc} - \frac{m^4}{M^2} + M^2 x^8 -... & 0 \\
    0 & M^2 + M^2 x^4 - ...\end{array}
\right ),
\end{eqnarray}
the expansion parameter of the rotation matrix (\ref{rotation}) is
only of the order $x$ and the corrections are thus always very small
as long as $\Lambda_{GUT} \gg m$. If we take $\Lambda_{GUT}=1000$ GeV
and $m=266$ GeV, one finds that the corrections to the Higgs boson
squared mass are small and that it is still negative. Our model in
that framework provides a natural mechanism to trigger the Higgs
mechanism. The four heavy degrees of freedom would have masses in the
TeV range. If the grand unification scale is in the TeV region and if
the second boson $H$ has Yukawa type couplings with the standard model
fermions, we expect that flavor changing neutral current transitions
should take place at an observable rate.

We should like to conclude by emphasizing that the main achievement of
the mechanism proposed in this Letter is to induce the Higgs mechanism
via a seesaw mechanism. If the seesaw-Higgs mechanism is implemented
in a grand unified theory, the hierarchy problems is solved. An
intermediate energy scale in the $10^9$ GeV energy region is
predicted. Depending on the details of the breaking of the unification
group, the naturalness problem could be solved using the scale
invariance argument presented in \cite{Bardeen:1995kv}. This
issue is relevant to model building and will be considered in a
forthcoming paper.

\section*{Acknowledgements}
 The author is grateful to H.  Fritzsch, L. B. Okun and Z. Xing for
  enlightening discussions and for their valuable comments concerning
  the draft of this Letter. Furthermore he would like to thank G.
  Buchalla, A. Hebecker, S.  Heinemeyer, A. Leike for useful
  discussions.


\begin{thebibliography}{10}  
\bibitem{Glashow}
S.~L.~Glashow,
Nucl.\ Phys.\  {\bf 22} 579 (1961);
S.~Weinberg,
Phys.\ Rev.\ Lett.\  {\bf 19} 1264 (1967);
A.~Salam and J.~C.~Ward,
Phys.\ Lett.\  {\bf 13}, 168 (1964).

\bibitem{Higgs:1964pj}
P.~W.~Higgs,
Phys.\ Lett.\  {\bf 12} (1964) 132; 
Phys.\ Rev.\ Lett.\  {\bf 13} (1964) 508; 
Phys.\ Rev.\  {\bf 145} (1966) 1156; 
F.~Englert and R.~Brout,
Phys.\ Rev.\ Lett.\  {\bf 13} 321 (1964); 
G.~S.~Guralnik, C.~R.~Hagen and T.~W.~Kibble,
Phys.\ Rev.\ Lett.\  {\bf 13} 585 (1964); 
T.~W.~Kibble;
Phys.\ Rev.\  {\bf 155}, 1554 (1967).


\bibitem{Georgi:1974sy}
H.~Georgi and S.~L.~Glashow,
Phys.\ Rev.\ Lett.\  {\bf 32}, 438 (1974).

\bibitem{Fritzsch:1975nn}
H.~Fritzsch and P.~Minkowski,
Annals Phys.\  {\bf 93}, 193 (1975),
H.~Georgi, in {\it Particles and Fields}, (AIP, New York, 1975).

\bibitem{'tHooft:rn} G.~'t Hooft,
Nucl.\ Phys.\ B {\bf 35}, 167 (1971);
Nucl.\ Phys.\ B {\bf 33}, 173 (1971);
G.~'t Hooft and M.~J.~Veltman,
Nucl.\ Phys.\ B {\bf 44}, 189 (1972);
Nucl.\ Phys.\ B {\bf 50}, 318 (1972).


\bibitem{'tHooft:1980xbis} G.~'t~Hooft, in ``Recent Developments In
  Gauge Theories'', Carges\`e 1979, ed.  G.~'t~Hooft et al. Plenum
  Press, New York, 1980, Lecture III, p.135, 
L.~Susskind,
Phys.\ Rev.\ D {\bf 20}, 2619 (1979).

\bibitem{Hill:2002ap}
C.~T.~Hill and E.~H.~Simmons,
arXiv:hep-ph/0203079.

\bibitem{Haber:1984rc}
H.~E.~Haber and G.~L.~Kane,
Phys.\ Rept.\  {\bf 117}, 75 (1985).

\bibitem{Arkani-Hamed:1998rs}
N.~Arkani-Hamed, S.~Dimopoulos and G.~R.~Dvali,
Phys.\ Lett.\ B {\bf 429}, 263 (1998)
[arXiv:hep-ph/9803315];
L.~Randall and R.~Sundrum,
Phys.\ Rev.\ Lett.\  {\bf 83}, 3370 (1999)
[arXiv:hep-ph/9905221].


\bibitem{Atwood:1996vj}
D.~Atwood, L.~Reina and A.~Soni,
Phys.\ Rev.\ D {\bf 55}, 3156 (1997)
[arXiv:hep-ph/9609279].


\bibitem{Coleman:jx}
S.~R.~Coleman and E.~Weinberg,
Phys.\ Rev.\ D {\bf 7}, 1888 (1973).

\bibitem{Gunion:1989we} J.~F.~Gunion, H.~E.~Haber, G.~L.~Kane and
  S.~Dawson, ``The Higgs Hunter's Guide,'' SCIPP-89/13.



\bibitem{Fritzsch:1999ee}
H.~Fritzsch and Z.~z.~Xing,
Prog.\ Part.\ Nucl.\ Phys.\  {\bf 45}, 1 (2000)
[arXiv:hep-ph/9912358];
R.~N.~Mohapatra,
arXiv:hep-ph/9910365.

\bibitem{Fritzsch:1979zq}
H.~Fritzsch,
Nucl.\ Phys.\ B {\bf 155}, 189 (1979).

\bibitem{Gell-Mann:vs} M.~Gell-Mann, P.~Ramond and R.~Slansky, in {\it
    Supergravity}, P. van Nieuwenhuizen \& D.Z. Freedman (eds.), North
  Holland Publ. Co., 1979;
T. Yanagida, in {\it Proceedings of the Workshop on Unified Theory and
  the Baryon Number of the Universe}, KEK, Japan, 1979.


\bibitem{Dobrescu:1997nm}
B.~A.~Dobrescu and C.~T.~Hill,
Phys.\ Rev.\ Lett.\  {\bf 81}, 2634 (1998)
[arXiv:hep-ph/9712319];
R.~S.~Chivukula, B.~A.~Dobrescu, H.~Georgi and C.~T.~Hill,
Phys.\ Rev.\ D {\bf 59}, 075003 (1999)
[arXiv:hep-ph/9809470];
H.~J.~He, C.~T.~Hill and T.~M.~Tait,
Phys.\ Rev.\ D {\bf 65}, 055006 (2002)
[arXiv:hep-ph/0108041].

\bibitem{Ito:2000cj}
M.~Ito,
Prog.\ Theor.\ Phys.\  {\bf 106}, 577 (2001)
[arXiv:hep-ph/0011004].



\bibitem{Mohapatra:1992dx}
R.~N.~Mohapatra and M.~K.~Parida,
Phys.\ Rev.\ D {\bf 47}, 264 (1993)
[arXiv:hep-ph/9204234].


\bibitem{Pati:1974yy}
J.~C.~Pati and A.~Salam,
Phys.\ Rev.\ D {\bf 10}, 275 (1974).

\bibitem{Bardeen:1989ds}
Y.~Nambu and G.~Jona-Lasinio,
Phys.\ Rev.\  {\bf 122}, 345 (1961);
W.~A.~Bardeen, C.~T.~Hill and M.~Lindner,
Phys.\ Rev.\ D {\bf 41}, 1647 (1990).

\bibitem{Kaplan:1983fs}
D.~B.~Kaplan and H.~Georgi,
Phys.\ Lett.\ B {\bf 136}, 183 (1984).


\bibitem{Calmet:2000th}
X.~Calmet and H.~Fritzsch,
Phys.\ Lett.\ B {\bf 496}, 161 (2000)
[arXiv:hep-ph/0008243];
Phys.\ Lett.\ B {\bf 525}, 297 (2002)
[arXiv:hep-ph/0107085];
X.~Calmet,
Phys.\ Lett.\ B {\bf 510}, 221 (2001)
[arXiv:hep-th/0008189];
Ph.D. Thesis, {\it A Duality as a Theory for the Electroweak
  Interactions}, Ludwig Maximilian's University (2002).




\bibitem{Bardeen:1995kv}
W.~A.~Bardeen,
FERMILAB-CONF-95-391-T, {\it Presented at the 1995 On scale symmetry
of the action at the quantum level. Ontake Summer Institute, Ontake
Mountain, Japan, Aug 27 - Sep 2, 1995}.


\end{thebibliography}
\end{document}